\documentstyle[11pt,epsf]{article}

\input psfig
\def\beq{\begin{eqnarray}}
\def\eeq{\end{eqnarray}}
\def\bea{\begin{eqnarray*}}
\def\eea{\end{eqnarray*}}

\def\centeron#1#2{{\setbox0=\hbox{#1}\setbox1=\hbox{#2}\ifdim
\wd1>\wd0\kern.5\wd1\kern-.5\wd0\fi
\copy0\kern-.5\wd0\kern-.5\wd1\copy1\ifdim\wd0>\wd1
\kern.5\wd0\kern-.5\wd1\fi}}
\def\ltap{\;\centeron{\raise.35ex\hbox{$<$}}{\lower.65ex\hbox{$\sim$}}\;}
\def\gtap{\;\centeron{\raise.35ex\hbox{$>$}}{\lower.65ex\hbox{$\sim$}}\;}

\def\singleandthirdspaced{\baselineskip=\normalbaselineskip\multiply
    \baselineskip by 130\divide\baselineskip by 100}
\def\singlespaced{\baselineskip=\normalbaselineskip}

\newcommand{\newc}{\newcommand}
\newc{\qbar}{{\overline q}}
\newc{\Kahler}{K\"ahler }
\newc{\deltaGS}{\delta_{\rm GS}}
\begin{document}
\begin{titlepage}
\begin{flushright}
{\large
hep-th/9903212 \\
SCIPP-99/02\\

}
\end{flushright}

\vskip 1.2cm

\begin{center}

{\LARGE\bf Seeking the Ground State of String
Theory}\footnote{Invited talk at the Yukawa-Nishinomiya Symposium,
Nishinomiya City, November 1998.}

\vskip 1.4cm

{\large
Michael Dine}
\\
\vskip 0.4cm
%{\it $^a$Stanford Linear Accelerator Center,
%     Stanford CA 94309} \\
{\it Santa Cruz Institute for Particle Physics,
     Santa Cruz CA 95064  } \\
%{\it $^c$Physics Department,
%     University of California,
%     Santa Cruz CA 95064  } \\

\vskip 4pt

\vskip 1.5cm

\begin{abstract}

The greatest obstacle to developing a string phenomenology is our
lack of understanding of the ground state. We explain why the
dynamics which determines this state is not likely to be
accessible to any systematic approximation. We note that the
racetrack scheme, often cited as a counterexample, suffers from
similar difficulties. We stress that the weakness of the gauge
couplings, the gauge hierarchy, and coupling unification suggest
that it may be possible to extract some information in a
systematic approximation.  We review the ideas of Kahler
stabilization, an attempt to reconcile these facts. We consider
whether the system is likely to sit at extremes of the moduli
space, as in recent proposals for a low string scale. Finally we
discuss the idea of Maximally Enhanced Symmetry, a hypothesis
which is technically natural, compatible with basic facts about
cosmology, and potentially predictive.
\end{abstract}

\end{center}

\vskip 1.0 cm

\end{titlepage}
\setcounter{footnote}{0}
\setcounter{page}{2}
\setcounter{section}{0}
\setcounter{subsection}{0}
\setcounter{subsubsection}{0}

%%%%%%%%%%%%%%%%%%%%%%%%%%%%%%%%%%%%%%%%%%%
%%%%%%%%%%%%%%%%%%%%%%%%%%%%
\singleandthirdspaced

%\begin{document}

\section{Introduction}

In thinking of Yukawa's great work, one cannot help but consider
the importance of asking the right questions.  It is not merely
that Yukawa predicted the existence of a particular particle, but
that the questions which he asked, and the answer which he
provided, remain fruitful
to this day.  In thinking about the
fundamental interactions, we would like to pose and answer
similar, qualitative questions.  This has been the spirit of this
conference.  It would be presumptuous to suppose that the
questions which I will pose here will be of such importance, or
the answers so significant.    It is likely, as we will see, that
it is premature to ask these questions.  It is also possible that
the questions I pose here will someday seem inappropriate, in much
the way we no longer see the computation of the total cross
section as an important problem in strong interactions.  Still, I
hope that the questions which I will phrase here will be helpful
in confronting some of the issues which we face in thinking about
the formulation of string theory and its connection with nature.

Up until now, our approach to string theory has suffered
from a
certain schizophrenia. At weak coupling, we have a beautiful
picture, containing gravity, gauge interactions, chirality,
generations, calculable interactions, and other features which we
view as crucial to any fundamental theory.  But there are also
problems with this picture.  First, there are many vacua.  These
vacua carry both discrete and continuous labels.  The discrete
labels, especially the number of supersymmetries, are extremely
useful in gaining control over the theory; they can be used to
severely constrain, for example, the effective lagrangian for the
moduli, the fields whose expectation values correspond to the
continuous labels.  The more supersymmetry, the stronger the
constraints and the more powerful the statements we can make.
This power is the origin of virtually all of the recent
developments connected with duality. Unfortunately, one of the
easiest statements to make about ground states with more than four
supersymmetries ($N>1$ in four dimensional counting) is that they
{\it are} ground states, perturbatively and non-perturbatively.

For the interesting cases with $N \le 1$ supersymmetry, there is
no such simple argument, in general.   Generically, perturbative ground
states with four or less supersymmetries are
unstable.  Assuming that string theory describes nature, duality
is not likely to be of much help in establishing the properties of
the non-supersymmetric state which corresponds
to the world we observe. Duality generally relates very strong coupling
in one theory to weak coupling in another theory. Yet a simple
argument shows that any stable ground state of the theory is
either degenerate with other ground states or lies at a point in
the moduli space where no weakly coupled description is
possible\cite{dineseiberg}. While this argument was originally
formulated with the heterotic string in mind, it applies to {\it
any} weak coupling description.  The point is simply that at weak
coupling, any potential for the modulus which describes the
coupling necessarily goes to zero.Recent developments in duality
are thus not directly useful for addressing the problem of vacuum
stability\cite{dineshirman}.  If a very strong coupling region of
one theory corresponds to a weakly coupled description of another,
than it suffers from the same problem. This suggests that in
seeking the phenomenologically interesting ground state(s) of the
theory, we need to look for regions of the moduli space which
admit no weak coupling description.\footnote{The notion of a
moduli space, in this context, is a bit fuzzy; we have in mind
the {\it classical} moduli space;
later, we will introduce the notion of an approximate moduli
space.}  David Gross has dubbed this the ``Principal of Minimal
Calculability (PMC).''

Still, nature exhibits weak coupling and seems to exhibit
perturbative unification. Surely, these are important clues.  One
of the themes of this lecture will be an effort to reconcile these
seemingly contradictory facts.  A related fact, which also figures
heavily in this talk, is that nature exhibits hierarchies.  From
the perspective of string theory, the issue is:  how does a theory
with no parameters generate large (small) pure numbers.  We will
confront these issues quite directly in this discussion.  We will
see that the traditional particle physics answers provide only
limited comfort.  For example, we often say that hierarchies can
be understood through the slow running of couplings, or through
the natural appearance in field theories of expressions involving
$e^{-8 \pi^2 /g^2}$.  We say that it is natural that $g$ should be
of order $1$, and therefore the exponential can be extremely
small. But, especially in light of duality, a more natural
condition is that $g^2 \over 4 \pi$ should be $1$.  Then the
exponential should not be thought of as small at all.  To my
knowledge, there are only three proposals in the literature to
deal with this conundrum, and I will review them here.  The first
of these is known as the ``racetrack" scheme\cite{racetrack}. The
idea is that one has several gaugino condensates (or similar
sources of moduli superpotentials). It is assumed that the
superpotential is a sum of the superpotentials generated by gluino
condensation in each sector, i.e. it has the form \beq W= \sum
e^{-{8 \pi \over b_i g_i^2}}. \eeq Generically, any ground states
one finds in such a picture will suffer from the problem described
above, that $e^{-8 \pi^2/g^2}$ is not small, since the various
terms in the sum must be comparable. It has been argued, however,
that for some choices of low energy gauge groups, large
hierarchies can result if there are terms in the sum with large
$\beta$-functions.  This proposal has been most persuasively
developed by Kaplunovsky and Louis\cite{racetrack}. We will see
that while this is a logical possibility, it cannot be studied in
a systematic approximation, and it seems unlikely that minima with
the desired properties exist. The second proposal is known as
Kahler stabilization, and will be briefly reviewed
here\cite{coping}.  We will spend most of our time, however, on a
third proposal, that of ``Maximally Enhanced Symmetry"\cite{dns}.
This proposal, as we will see, has problems of its own, but if we
hypothesize that it is correct, it makes definite predictions for
low energy physics, and also suggests an approach to developing a
real string phenomenology.

\section{Fixing the Moduli}

\subsection{Generalities}

String theory is a theory without free parameters, but the role
of
parameters is played by the moduli.  Even without any detailed
understanding of string dynamics, there are only a few logical
possibilities for the fate of the moduli:
\begin{itemize}
\item  The moduli are not fixed.  Perturbatively and
non-perturbatively, their potential vanishes.  This is
the case for $N>1$ supersymmetry in four dimensions and for supersymmetry
in $D>4$.  In such cases, the
supersymmetry prevents one from writing any potential for the moduli in the
low energy effective theory at
all.
\item  The moduli are unstable to
runaway to $\infty$.  Consider, for example, the modulus which
describes the string coupling.  One expects that any potential
which is generated either perturbatively or non-perturbatively
will vanish as the coupling tends to zero.  One can imagine
loopholes to this argument, but it is certainly true of all known
cases. A similar argument applies to moduli which describe, for
example, compactification radii, at least in cases with an
approximate, low energy supersymmetry\cite{bdn}.  As the radii
become large, the theory becomes a theory in more that four
dimensions with supersymmetry, and in such theories the energy
necessarily vanishes. Again, one can conceive of loopholes to this
argument, but it holds for all known cases.
\item  The moduli are fixed.  Necessarily this occurs for couplings of order one.
This follows from our
discussion above:  if any weak coupling approximation is valid, the potential
cannot have a minimum.  A
similar argument holds for the moduli which describe compactification (at least,
as discussed above, in cases with low energy supersymmetry).  Of
course, we can hope for accidents.  Ratios of scales might be small, couplings in
effective lagrangians
might be small, simply by accident.  But the underlying theory is still likely to be
strongly coupled.
Without some further assumptions, one has little hope of predicting anything under
theses circumstances.  As we will discuss below, supersymmetry, holomorphy, and
other symmetries {\it may} allow one to make
some predictions.
\end{itemize}

String duality is an exciting development, but it doesn't help directly with this
problem.  Indeed, these
arguments imply that we expect realistic ground states of string theory to lie
where {\it no} weak coupling
description is valid (this is the PMC alluded to above).

How, then, can we imagine developing any phenomenology?  How can
we reconcile these remarks with the facts that the gauge couplings
we observe in nature are small and seem to unify? In this lecture,
I will review what has been said about these questions, and offer
some speculations.  I will focus on the third possibility in our
list above, that the moduli are fixed and in some sense of order
one. I will focus principally on the questions:  why is $\alpha
\ll 1$, and why is ${m_W \over m_P} \ll 1$.  I won't offer much
insight into the questions:  what is the origin of the fermion
mass hierarchy, and, most importantly, why is the cosmological
constant so small.  It will be clear from the discussion that our
lack of an answer to this last question is an extremely serious
limitation, and quite possibly an indication that all of this
discussion is premature.

Before turning to general theoretical issues, I would like to
mention one more possible clue and constraint. This is the
cosmological moduli problem\cite{bkn}.  Most proposals for
supersymmetry breaking in string theory postulate that the moduli
develop a shallow potential, typically with a minimum at some more
or less random value.  In that case, the early universe has no
reason to start out close to the ground state and the system
generically stores too much energy.  The conventional solution to
this problem is to suppose that when these moduli decay, they heat
the universe above nucleosynthesis temperatures, and produce the
observed baryon asymmetry at the same time. This requires that the
moduli be quite heavy\cite{bkn}. An alternative possibility is
that all of the moduli are charged under symmetries at the
minimum, so that the values of these fields in the early universe
can naturally coincide with those at the present time\cite{drt}.
Prior to the recent developments in string duality, it was not
possible to say much about this possibility. Now it is easy to
construct examples of this phenomenon. This will be the focus of
the latter part of this talk.

It is worthwhile to first review the conventional particle physics
wisdom about small numbers.  Usually we say that the gauge
couplings ($g_s$, $g$, $g^{\prime}$) are numbers of order $1$.
Thus it is natural that $\alpha_{gut} \sim {1 \over 30}$, for
example.  This also provides a way of understanding hierarchies,
since non-perturbative effects in weakly coupled field theories
typically behave as $e^{-{8 \pi^2 \over g^2 b}}$.  But this begs
the question:  why is it $g$ which is ${\cal O}(1)$, and not
$\alpha$.  In the case of electric-magnetic duality, for example,
$\alpha=1$ at the self-dual point.  We will ¦confront this
issue shortly.  About small Yukawa couplings, there are a few
ideas.  Most assume that there are approximate symmetries, broken
by the expectation values of fields. In string theory, it has long
been noted that some Yukawa couplings vanish exponentially with
compactification radius, and this could be a source of Yukawa
hierarchies\cite{newissues}.  A version of this idea suggested by
brane physics has recently been studied by
\cite{dimopoulosflavor,dvaliflavor}. About the cosmological
constant, there is no real conventional wisdom, and I won't have
anything to add today. Recently, Kachru and Silverstein\cite{ks}
have exhibited non-supersymmetric models in which the cosmological
constant vanishes, and Harvey\cite{harvey} has noted that in some
string theories the cosmological constant may be exponentially
small. Witten earlier made an interesting proposal motivated by
string dualities\cite{wittencosmo}. But none of these ideas is as
yet complete.

It is also useful to recall the conventional approaches to string
phenomenology. Essentially all string phenomenology ignores the
problem of moduli and simply assumes that the moduli are fixed to
some convenient values.  The couplings are invariably assumed
weak.  The rationale is that, after all, weakly coupled strings
look much like nature.  Prior to 1995, virtually all string model
building started with the assumption that only the heterotic
string was phenomenologically viable.  In this theory, the unified
coupling is related to the string coupling and the
compactification volume through \beq \alpha_{gut}\approx {g_s^2
\over V} \eeq Here $g_s$ is the dimensionless string coupling and
$V$ is the compactification volume in units of the string tension.
Requiring that $g_s \le 1$, so that a weak coupling string
description should be valid, yields that $V \sim 1$, while $M_s
\sim M_{gut} \sim M_p$\cite{vadimscales,dsscales}.  The
requirement that $g_s < 1$, however, was always artificial, a
reflection of our wishful thinking that the theory should be
weakly coupled. Our post 1995 understanding of duality permits
many new possibilities. Perhaps the simplest of these is the
proposal by Witten\cite{wittencompact}, following the realization
of Horava and Witten\cite{horavawitten}
 that the strongly coupled
heterotic string theory is in fact an eleven dimensional theory
with two walls in the eleventh dimension. Taking the values of the
gauge coupling and unification scale at face value, this could be
the appropriate description of the theory. It leads to a picture
in which the world is approximately five dimensional; six
dimensions are compactified, say, on a Calabi-Yau space with
radius of order one in eleven-dimensional Planck units, while one
dimension is significantly larger. More radical
possibilities\cite{lykken,dimopoulos,precursors,otherextradim}
have been proposed recently, and will be discussed further below.
One intriguing possibility is that the standard model lives on a
brane.  In this case, the gauge couplings are independent of the
compactification volume, and one can, again, consider the
possibility that the compact space is large. In all of these
pictures, one must still ask why some dimensions are large.

Finally, what {\it is}
the conventional wisdom about fixing the moduli?  There are a number
of approaches:
\begin{itemize}
\item  Ignore the issue.
\item  Assume that the moduli are stabilized at strong coupling, and that the
smallness of the observed couplings is an accident.  This could be
well be the answer, and is in line with the PMC enunciated
earlier. But it is very disappointing, and leaves us without an
explanation of why string theory gets even qualitative things
right.
\item  Racetrack models\cite{racetrack}:
In theories with two or more gaugino condensates, it has been argued
that there is a superpotential for the moduli of the form \beq W=
\sum C_a(T) e^{-{8 \pi \over b_a \alpha_a(T)}}. \label{wracetrack}
\eeq Then one can have, for example, isolated supersymmetric
minima with fixed, large values of the moduli, provided that some of the $b_a$
are large and nearly equal. Supersymmetry can then be broken at a lower scale.
\item  Kahler stabilization\cite{coping}:
In this picture, holomorphic quantities,
such as $e^{-8 \pi^2 \over g^2}$
and the holomorphic gauge couplings are small.  However,
non-holomorphic
quantities, such as the Kahler potential, are
assumed to receive large corrections, and to be responsible for
the stabilization of the moduli.
\item  Large Topological charges:  The authors of \cite{dimopoulosmoduli}
argue that large topological charges could stabilize compactification
radii at large values.  One could imagine that similar effects stabilize
other moduli.  The question would then be whether why these charges
take such values.
\item  Maximally Enhanced Symmetry\cite{dns}:  This is the natural
postulate that the ground state lies at a point where all of
the
moduli transform under unbroken symmetries.  This postulate
makes some definite predictions.  Whether there really exist
string vacua with such symmetries, and simultaneously with small
effective gauge couplings, is an open question.
\end{itemize}

We don't have much to add to  the first two items.  In the next
section, we will explain why the multiple gaugino condensate idea
cannot be studied systematically. The following section then explains
the basic ideas of Kahler stabilization. The final section is
devoted to the hypothesis of maximally enhanced symmetries.

\subsection{Racetrack Models}

Racetrack models have been offered to explain how string vacua
might arise at perturbatively weak couplings.  The idea is that
the superpotential of the theory is a sum of terms of the form of
[\ref{wracetrack}].  In fact, as has been explained in
\cite{racetrack}, it is quite natural that effects associated with
low energy dynamics be larger than stringy non-perturbative
dynamics.  Noting that low energy gauge groups in string theory
can be extremely large, the authors of \cite{racetrack} argue that
if two groups have very large $\beta$-functions, some of the
moduli can be fixed, without breaking supersymmetry.

There are several difficulties with this picture.  All are related
to the problem that that there is no small parameter to justify
the approximations; i.e. there is nothing like the $N$ of the
large $N$ expansion\cite{seibergpc}. In this version of the
mechanism, for example, $x=e^{-8 \pi^2 b\over g^2}$ should be of
order one.  But this means that the scale of the low energy groups
is of the order of the fundamental scale, so the low energy
analysis is not really consistent.  Alternatively, higher order
operators complicate the the analysis.  As a result, it is
difficult to determine whether a vacuum state even exists.  Other
versions of the scheme suffer from similar difficulties.  One can
obtain smaller $x$ at the price of fine tuning, but it is still
difficult to obtain a small cosmological constant.  In all of
these versions, the Kahler potential is not calculable; this is
a particular important issue in versions of the scheme in which
supersymmetry is broken by the condensates. It would seem that one
should have simply hypothesized the desired result:  the coupling
is fixed in a way that the gauge coupling is small.  One has no
control over the final answer.  These issues will be explored in a
future publication.

\subsection{Kahler Stabilization}

Quite generally, if the superpotential is responsible for
stabilizing the moduli, it is unlikely that the effective
couplings can be weak, in the sense that $e^{-8 \pi^2/g^2}$ is
small.  This follows simply from holomorphy.  Consider, for
example, weakly coupled string theory.  We expect that the
superpotential is roughly of the form \beq W=e^{-S} + b e^{-2S} +
\dots \eeq If the Kahler potential is not significantly modified
from its tree level form, then
at any minimum of the potential,
$e^{-S} \sim e^{-2S} \sim 1$.  Thus the coupling is strong and
there is no hierarchy.\footnote{One might hope to get around this
by supposing that, say, considering, as in the racetrack schemes,
two terms, $ae^{-\alpha S}+ be^{-\beta S}$, and hoping to find a
minimum where $e^{\alpha-\beta}= b/a$, and $\alpha-\beta$ is
small, while $b/a$ is also small.  This is similar to the failed
racetrack schemes described above, but perhaps occurs in some
other context.}

We can be more precise if we exploit holormorphy and the discrete
axion shift symmetry. These restrict the couplings of $S$ to the
gauge fields to \beq ({1 \over 2
\pi^2 S} + e^{-S} + \dots)W_{\alpha}^2, \eeq while the
superpotential has the form\cite{coping}
 \beq W=e^{-S} + e^{-2S} +
\dots \eeq The Kahler potential, on the other hand, is not
restricted by holomorphy.  It is known that string perturbation
theory is not as convergent as field theory perturbation theory.
Assume, then, that there are large corrections even for $\alpha
\sim {1 \over 30}$. The full potential is given, in terms of $W$
and $K$ by (for simplicity, considering only the field $S$) \beq
V=e^{-K} \left [ \vert {\partial W \over \partial S} + {\partial K
\over \partial S} W \vert^2 \left ( {\partial^2 K \over \partial S
\partial S^*}
 \right )^{-1} - 3 \vert W \vert^2 \right ].
\label{vsugra}
\eeq

Kahler stabilization is the suggestion that this potential has its
minimum when $e^{-S}$ is small due to the structure of  $K$.   One
can certainly postulate forms for $K$ which yield a local minimum
of the potential for such values of $S$, with vanishing
cosmological constant (if one allows sufficient fine tuning). This
approach is predictive.  Because $e^{-S}$ is small, it predicts
coupling constant unification and that the
superpotential is not significantly altered from its weak coupling
form.  It also predicts that there is approximate, low energy
supersymmetry.  On the other hand, explaining, say, squark
degeneracy requires additional inputs.  While squark masses, for
example, are sometimes degenerate in the weak coupling limit, our
basic assumption is that the Kahler potential is very different
from its weak coupling form.  So one needs to
postulate, say, approximate flavor symmetries.

In this picture, other moduli are also fixed by the form of $K$.
Just as one does not expect the gauge couplings to be extremely small,
one does not expect large hierarchies of compactification radii.
This follows from the fact that as the radii become large, the theory
is effectively a supersymmetric theory in a higher dimension, where
one cannot write down a potential for the moduli.  In other words,
the potentials for the moduli which describe the size of the internal
dimension necessarily vanish as the size tends to infinity.

One doesn't expect, in such a picture, particularly large
hierarchies of compactification scales and $M_p$.   Of course, one
has provided no explanation of the cosmological constant puzzle.

\section{New Insights From Duality}

\subsection{Horava-Witten:  A Large Eleventh Dimension}

Duality has opened up new ways to think about these problems.  One
puzzle in the early days of superstring compactifications was
reconciling the observed values of the unified scale and couplings
with weakly coupled string theory.  In light of our understanding
of duality, it is reasonable, following Horava and
Witten\cite{horavawitten} and \cite{wittencompact} to suppose that
string theory is described by the heterotic string theory in a
strongly coupled regime.  In this regime, the theory looks eleven
dimensional with two walls in the $11$'th dimension separated by a
distance $R_{11}$\cite{wittencompact,bdscales}.  Calling $M_{11}$
the eleven dimensional Planck scale, and $V$ the compactification
volume of, say, some six dimensional Calabi-Yau manifold, one
finds \beq R_{11}^3 = {\alpha_{gut}^3 V \over 512 \pi^4 G_N^2}.
\eeq \beq M_{11}= R^{-1}(2 (4 \pi)^{-2/3} \alpha_{GUT})^{-1/6}.
\eeq In these equations, $G_N$ is the ordinary Newton constant.
Plugging in the observed values for the unification scale
($R^{-1}$) and the unified coupling constant, one finds \beq
M_{11} R_{11}= 72 ~~~~~~~~M_{11} R \approx 2. \eeq

In this picture, then, the eleven dimensional Planck scale is
close to the unification scale, while $R_{11}$ is significantly
larger.  This viewpoint has other interesting consequences.  For
example, it ameliorates the cosmological axion problem of string
theory\cite{bdscales,bdaxion}.  On the other hand, it is still
hard to understand the stabilization of the moduli.  For large
$R_{11}$, the bulk theory is approximately five dimensional.
Supersymmetry in five dimensions forbids a potential, so the
potential must tend to zero as $R_{11} \rightarrow \infty$.  This
can be made precise, by using the five dimensional supersymmetry
to restrict the form of $K$ and $W$, and one finds that the
potential does tend rapidly to zero, consistent with this
heuristic argument\cite{bdscales,ovrutetal}. Thus one expects
that, if there is a stable minimum of the potential, it occurs
when the various radii are of order $M_{11}$. The problem of
explaining why this ratio is of order $70$ seems similar to the
problem of understanding why the gauge coupling is of order
$1/30$.  Again, one needs something like Kahler stabilization of
the moduli.

\subsection{A More Radical Proposal:  String Theory at the $TeV$ Scale}

All of the previous discussion has been based on the idea that,
string theory being a theory without parameters, all dimensionless
couplings and ratios of scales should be numbers of order one
(with the exception of the supersymmetry breaking scale, which is
understood as the exponential of a number of order one). Indeed,
we have seen that in the supersymmetric case, one can prove
this.  Recently,
various authors have proposed that perhaps the string scale lies
at another familiar scale in physics, the scale of weak
interaction symmetry breaking\cite{lykken,dimopoulos}.  Such
ideas, in fact, had been considered in the past, but had not been
taken too seriously because such a possibility corresponds to {\it
enormous} string coupling, in the case of the heterotic string.
Newton's coupling is so small, or the Planck scale so large, in
such a picture, because the internal space has a very large
volume.  For example, if one compactifies the eleven dimensional
theory, one has \beq G_N = {{\rm TeV}^9 \over V^{(7)}}. \eeq In
particular, if all of the dimensions are of comparable size, than
$r$, the radius of the compact space, satisfies
$r \sim {\rm MeV}^{-1}$, while if, for example, two dimensions are
large, and the others are of order the fundamental scale, $r \sim
{\rm mm}$. Most of the proposals of this type assume that the fields of
the standard model live on a brane.  Gravity looks $4+n$
dimensional on scales small compared to $r$, where $n$ is the
number of large compact dimensions.  Exciting new phenomenological
possibilities exist: long range forces (in the mm case, and
possibly in others, as we will discuss below), production of large
numbers of Kaluza-Klein states, and production of stringy
excitations.

At first sight, this idea seems outrageous, but in fact it is
quite difficult to definitively rule out\cite{savasconstraints}.
There are several obvious problems to worry about:
\begin{itemize}
\item  Proton decay:  proton decay must be highly suppressed;
if the relevant scale is of order a $TeV$, then operators up to
very high dimension must be forbidden.  This can be arranged,
however, by assuming, for example, that there is a discrete
symmetry which is a large subgroup of baryon number.
\item  Other types of flavor violation:  These can
be suppressed if one assumes that the theory has a large flavor
symmetry, broken, perhaps, on distant
walls\cite{dimopoulosflavor,dvaliflavor}.  Still, these processes
constrain the scale to be greater than $5-10~ {\rm
TeV}$\cite{bkn}.
\item  Production of Kaluza-Klein modes:  In this picture,
typical Kaluza-Klein modes of the graviton couple with
gravitational strength.
However, there are a huge number of such
modes, so one needs to worry about processes in which one produces
these modes and they carry off energy.  The lower limits on the
string scale arising from these types of considerations are of
order a few $TeV$.
\item  Astrophysical constraints:  here one needs to worry about
production of these particles in red giants.  The problem is most
severe in the case of two compact dimensions.   Here one obtains
limits in the $ 30~ {\rm TeV }$ range if
$n=2$\cite{savasconstraints}; recently it has been argued that the
limit is $50~{\rm TeV}$\cite{scuyler}.  This means that it will be
difficult to observe the associated change in Newton's law, and is
certainly problematic from the perspective of the hierarchy
problem.
\item  Cosmology:  Even for general $n$
this is more problematic.  One has in these theories a serious
moduli problem, for example.  The authors of
\cite{savasconstraints} argue that, if the scale is not too low,
provided the universe was in the correct ground state shortly
before nucleosynthesis, production of Kaluza-Klein modes will not
spoil this.  As we will describe later, it is hard to imagine how
to establish such an initial condition.  One can contemplate, for
example, several stages of inflation, but one is still left with
a severe moduli problem\cite{dimopouloscosmology}.
\end{itemize}

While some of these issues may make one uncomfortable with the
idea of a low string scale, it is clear that these considerations
alone do not rule out the possibility.  The laboratory and
astrophysical
constraints at best place the lower limit on the
scale at $10~ {\rm TeV}$, and the cosmological constraints, while
potentially more severe, require assumptions about aspects of
early universe physics about which we don't have direct evidence.

Still, one can ask:  is there any physics which suggests a low
string scale.  The literature on this problem refers to the
hierarchy problem.  Indeed, if the scale is close to the weak
scale, then Higgs scalars with mass of this order are natural.
However, if the scale is $10 ~{\rm TeV}$, then this is less clear.
We have argued that the true ground state of string theory should
be strongly coupled.  But in this case, absent supersymmetry, one
expects any fundamental scalars to have masses of order the scale,
i.e. of order $10~ {\rm TeV}$.  So one has a fine tuning to at
least one part in several hundred, or perhaps even worse.   In
weakly coupled string theory, this might be acceptable.  At string
tree level, one often finds particles which are massless for no
symmetry reason.  Loop corrections to the mass might be in an
acceptable range.  However, as we have argued, it is not likely
that there is such a weak coupling parameter.

So there is already a potential hierarchy problem.  A more severe
problem arises when we ask:  how might we stabilize the radius at
such a large value?   After all, we have argued that most
dimensionless ratios in the theory should be of order $1$.  There
would seem to be two possible explanations for such large numbers.
One is that some modulus (not associated with the large
dimensions) takes an extreme value, and some largrangian parameter
relevant to fixing the size of the compact space is exponentially
small in this modulus.  It is not easy to see how this would work
in practice, and in any case fixing this modulus would represent
one more mystery.  An alternative possibility has been explored in
\cite{dimopoulosmoduli},following earlier suggestions of
Sundrum\cite{sundrum}; in this scenario, the large dimensions are
connected with the large value of some topological charge. The
problem of large radii is then replaced by the question of why
this topological charge is so large.

The problem of stabilization has been discussed in
\cite{dimopoulosmoduli} and further in \cite{bdn}.  In order to
discuss stabilization of the moduli, it is crucial to make some
assumptions about the way in which the cosmological constant is
cancelled.  One possibility is
that, independent of the value of the cosmological constant in the
effective low energy field theory, say at energy scales slightly
below the radius $r^{-1}$, the large distance cosmological
constant vanishes, for some unknown reason.  In this case, the
values of the bulk cosmological constant and the cosmological
constant of the brane theory are independent.  We
expect the brane cosmological constant to be of order $M^4$.  In
any large radius scenario, on the other hand, the bulk
cosmological constant must be {\it many} orders of magnitude
smaller than the value expected from dimensional analysis,
$M^{4+n}$.  Indeed, the bulk cosmological constant makes a
contribution to the masses of the Kaluza-Klein states of order
$\Lambda_b /M^{n+2}$.  This mass is greater than $1/R$ unless
$\Lambda_b < 1/(r^2 M^2) M^{4+n}$.  This is an additional fine
tuning which must be explained.  Moreover, the actual small value
of the number requires that there be some modulus besides the
radial dilaton which takes some extreme value. The authors of
\cite{dimopoulosmoduli} argue that such a value of $\Lambda_b$ is
at least plausible in the case that the bulk theory is
supersymmetric.  But in that case, one can show that the bulk
cosmological constant vanishes; a different analysis, described in
\cite{bdn}, is required.  I will return to this case below.
Finally, as noted in \cite{dimopoulosmoduli}, the required
toplogical charges are very large.

An approach which yields a more plausible picture is to assume
that the the cosmological constant must already nearly vanish in
the theory at scales below $r^{-1}$. More precisely, one assumes
that, on account of some unknown mechanism, the cosmological
constant in the bulk adjusts to cancel the contribution of the
brane, of possible higher curvature terms, and of any topological
charges in the internal dimensions.  The brane contribution is
expected to be of order $M^4$.  If one assumes that the internal
space is flat, then the radial dilaton mass is of order $\rm
{mm}^{-1}$, independent of the number of internal dimensions. This
is interesting, but, as explained in \cite{bdn}, a flat internal
space requires far more fine tuning than a curved one. Various
possibilities arise in the case of curved manifolds.  If the bulk
is not supersymmetric, the required topological charges are still
rather large, and generically the radial dilaton has mass of order
$1/R$. If the bulk theory is supersymmetric, the story is more
complicated, and depends sensitively on the values of $n$ and the
nature of the stabilizing charges.  One turns out to require
$n>4$, and even then, one typically finds that there are
contributions to the masses of the Kaluza-Klein states larger than
$1/r$.  If one is willing to suppose that supersymmetry is
hierarchically broken on the branes, one can find an example with
rather small charges and a radial dilaton light enough to affect
Cavendish experiments. This case comes closest to realizing a
solution of the hierarchy without very large parameters, and with
no more mystery than the usual one of understanding the smallness
of the observed cosmological constant.  The methods of \cite{bdn}
can also be applied to the possibility that the radii are large
due to some extreme values of other moduli.  Finally, in the case
$n=2$ (and certain suitable generalizations)
there may be additional possibilities.
If one has bulk supersymmetry,
the potential for $r$ is a function of $\ln(r)$.
If this function has a minimum for a value of $\ln(r) \sim 40$,
this would give rise to a large radius.  Moreover, because the
fields in the bulk vary logarithmically, and some inevitably
couple to $F \tilde F$, such a picture might account for the
smallness of the gauge couplings.\footnote{I thank Savas Dimopoulos,
Nima
Arkani-Hamed and John March-Russell
for discussions of this possibility.  Elaborations on
these ideas will appear elsewhere.  The special role
of $n=2$ has been discussed in
\cite{ab}.}  This picture of stabilization
has much in common with the idea of Kahler stabilization discussed
earlier.

Given a model for the stabilization of the radial dilaton, the
question of early universe cosmology is also brought into focus.
Perhaps the most serious issue is production of bulk modes.
As noted in \cite{savasconstraints}, if the temperature on the
brane is higher than some temperature, $T_o$, then the bulk modes
are overpopulated at nucleosynthesis.  In the case $n=2$, this temperature
is only a few MeV.  This already seems a fine tuning.  Moreover, in
this case, there may be other, very efficient, mechanisms for
production of bulk modes, as will be described below.
For any dimension, there is also a potential, very severe,
moduli problem.
In particular, if there is a period of inflation, it would seem
that the bulk cosmological constant, and hence the location of the
minimum, would be modified.  The authors of
\cite{savasconstraints} deal with this problem by supposing that
the inflaton lives on a brane. A specific proposal along these
lines was made in \cite{dvalitye}. This model, however,
illustrates several generic problems:  it is hard to obtain a
reasonable fluctuation spectrum, and it is hard to understand why
the reheating temperature is not of order $M$\cite{bdn}.  Other
potential problems abound\cite{morecosmo,lindekaloper,bdn}
Cosmology in higher dimensions opens many new and
interesting possibilities, but the difficulties look formidable.

All of this leads to the following conclusion:  given the current
state of our understanding, the view that the compactification
scales should be small is, at best, prejudice.  There is no
decisive theoretical argument that the fundamental scale of string
theory couldn't be a few ${\rm TeV}$.  It is important to keep
this possibility in mind, both in thinking about experiments and
in developing string phenomenology. On the other hand, the issues
raised above, and the absence of any compelling argument in favor
of large dimensions, in my view, provide some reinforcement for
the earlier prejudice.

\section{Maximal Symmetry}

We now turn to another approach to the problem of
moduli\cite{dns}, inspired, in part, by recent developments in
duality, and in part by the cosmological moduli problem\cite{bkn}.

Imagine the classical moduli space of some string
compactification.  If there are
points in this moduli space where
all of the moduli are charged under unbroken symmetries, then:
\begin{itemize}
\item   Such points are automatically stationary points of the full effective
action
\item   Because of their high degree of symmetry, it is natural for the early
universe to start out at such a point.
\end{itemize}

From studies of duality, we know that there are many such points
of ``Maximal Symmetry."  Probably the simplest example of this
phenomenon is provided by the IIB theory in $10$ dimensions.  This
has a well-known $SL(2,Z)$ symmetry, under which \beq \tau = {i
\over g} + a \eeq transforms as \beq \tau \rightarrow -{1 \over
\tau} ~~~~~~~\tau \rightarrow \tau+1. \eeq With $a=0$, the first
transformation has a self-dual point, a particular value
of the
coupling at which the $Z_2$ symmetry is ``restored."  At this
point, the dilaton transforms.

Upon compactification, one can construct many more examples of
this phenomenon, with varying amounts of supersymmetry.  For
example, one can consider toroidal compactification
of the IIB
theory, with special radii and angles for the torus; Gepner
compactifications of the Type II theory, and many others.  One
interesting example in four dimensions is provided by toroidal
compactification of the heterotic string.   If one takes the torus
to be a product of circles, each at the appropriate $SU(2)$ point,
then all of the moduli are charged under the $SU(2)$'s except the
dilaton.  But this theory also has an $SL(2,Z)$ symmetry; at the
self dual point, $S$ transforms.  Presumably this phenomenon
occurs also in theories with $N=1$ supersymmetry; this is
currently under study.

As we stated, such points are automatically stationary points of
the effective action and thus candidate minima.\footnote{In
theories with more than four supersymmetries, the moduli space is
generally exact quantum mechanically, i.e. there is no potential
for the moduli.  We have in mind with these remarks models with
four or less (zero) supersymmetries.}   Moreover, it is natural
for the early universe to favor states with high degrees of
symmetry, so this proposal  solves the cosmological moduli
problem.  However, there are some obvious objections to this
possibility.  In particular, one expects that, generically,
$\alpha={\cal O}(1)$ at these points.  This follows from our
discussion in the previous section regarding holomorphy.  It also
follows for some of our particular examples.  In the case of
electric-magnetic duality (the heterotic example above), the Dirac
quantization condition shows that the gauge coupling is indeed of
order $1$ at the self-dual point.   In the case of  four
supersymmetries, the situation might be better. First, as in the
discussion of the previous section, Kahler potential effects might
be relevant in the symmetry restoration phenomenon.  A toy example
of this was provided in \cite{dns}.  It would be of interest to
survey examples of the enhanced symmetry phenomenon to determine
if the couplings are ever small.

In any case, for the rest of this section, we will adopt a set of
hypotheses
and explore their consequences. In particular, we will
assume that, at some high scale, $M$, one has
\begin{itemize}
\item Maximally enhanced symmetry:  all
of the moduli transform under symmetries.  An interesting and
perhaps appealing
version of this hypothesis is that there are no moduli.
\item  Approximate $N=1$ supersymmetry.
\item  Small, unified gauge couplings.  In particular,
$e^{-S}$ is an extremely small number.
\end{itemize}

Before going further, we should note that gaugino condensation
cannot play a role in supersymmetry breaking at such points.  This
is because linear couplings of the moduli to the gauge
fields,
${\cal M} W_{\alpha}^2$, are forbidden by the symmetries.  This
means will mean that we require supersymmetry breaking at
relatively low energies.

\subsection{The Case of No Moduli}

Consider the possibility that there are no moduli.  This has two
immediate consequences. First, supersymmetry cannot be broken by
high energy string effects, but must be broken by effects which
are visible in the low energy theory.  To understand this,
note that any supersymmetry breaking
effect must be describable in terms of a superpotential which is a
function of the light fields, $W(\Phi)$.  In order to obtain
supersymmetry breaking, one needs a linear term for some singlet
field.  But it is not plausible that there are such light fields
at strong coupling.\footnote{Here, we are assuming that some sort
of conventional notion of naturalness holds in strongly coupled
string theory, i.e. when one includes both perturbative and
non-perturbative effects. Note we are also ignoring the effects of
field-independent constants in the superpotential. This follows
from our basic assumption that supersymmetry is not broken at the
high energy scale.  It can be enforced by an unbroken discrete R
symmetry.}

This observation, in turn, has the consequence that whatever
breaks supersymmetry must be visible in the low energy dynamics.
This is likely to mean that there is a supersymmetry breaking
hidden sector.  What is the scale?  If there are no singlets,
couplings of the form $\Phi W_{\alpha}^2$ cannot be the source of
gaugino mass, as is usually assumed.\footnote{Several authors have
noted recently that at one loop, effects associated with the
Kahler anomaly can lead to gaugino masses.  However, these effects
tend to be quite small, and to require a large scale for
supersymmetry
breaking\cite{murayamakahler,randallkahler,baggerprivate}.}  This
suggests that supersymmetry breaking must be a low energy
phenomenon, presumably mediated by gauge interactions. In other
words, this general framework predicts something like
low energy gauge mediation.

\subsection{Maximal Symmetry}

Now consider the case that there are moduli, and that the minimum
of the potential lies at a point of maximal symmetry.  This turns
out to be similar to the case of no moduli, in its low energy
consequences.  Supersymmetry breaking again must be a low energy
phenomenon, since the symmetries forbid terms linear in the
fields.  Gaugino condensation and its generalizations are also
forbidden, since couplings such as $\Phi W_{\alpha}^2$ do not
respect the symmetries.  So, just as in the case of no moduli,
supersymmetry breaking must be a low energy phenomenon, presumably
mediated by gauge interactions.

One question which we can ask in this framework is:  how are the
moduli stabilized?  There are several rather natural
possibilities. Perhaps the most interesting is the following.
Suppose that all of the moduli are charged under standard model
gauge symmetries. This is not such an outlandish suggestion.  The MSSM has
approximate flat directions in which all the gauge symmetry is
broken.  For example, there is a direction parameterized by $Q Q Q
L$, i.e. \beq Q= \left ( \matrix{v & 0 & 0 \cr 0 & v & 0 \cr 0 & 0
& v} \right )  ~~~~~~ L = (0,v) \eeq where the $Q$ field is
written as a matrix in color and flavor.  This flat direction has
$12$ parameters, or 12 candidate moduli.  Note that this direction
can be exactly flat even without string miracles; a discrete $R$
symmetry under which $Q$ and $L$ are neutral can insure this.

Now suppose supersymmetry is broken in a hidden sector, with the
breaking communicated by gauge fields as in usual models of gauge
mediation.  Then the fields $Q$ and $L$ will receive positive
masses-squared from low energy loop effects.  This means that the
potential has a minimum at the symmetric point.

\section{The Cosmological Constant Problem}

Within each of these hypotheses, the cosmological constant problem
remains a significant puzzle.  Within the framework of Kahler
stabilization, we had to suppose a fine tuning of the Kahler
potential to obtain vanishing $\Lambda$.  In the case of Kahler
stabilization, we assumed that the supersymmetry was broken at the
intermediate scale, and transmitted to light fields by effects
suppressed by $1/M_p$.   Examining eqn. \ref{vsugra}, we see that
the potential possesses terms of opposite signs.   For
intermediate scale supersymmetry breaking, these terms are of the
same order of magnitude.  If the Kahler potential has just the
right form (i.e. if it is tuned in precisely the right way), these
terms can cancel. In the case of maximal symmetry, however, we
have argued that the breaking must be at a low scale.  In this
case, the contributions to the $3 \vert W \vert^2$ term from the
susy-breaking dynamics are suppressed by powers of the breaking
scale over $M_p$. Something more is needed if the cosmological
constant is to vanish at the level of the effective lagrangian. It
is necessary that there be a large constant in the superpotential
in order to cancel the contribution to the vacuum energy coming
from low energy supersymmetry breaking.  This constant could be
generated by gluino condensation in a pure gauge theory. In the
absence of moduli, such condensation does generate a constant in
$W$.  This constant would have to be just of the right size to
cancel the cosmological term from the other sectors.  This is
arguably a troubling feature of gauge mediation in general.  Of
course, given our total lack of understanding of the cosmological
constant problem, perhaps this concern is misplaced.

\section{Conclusions}

If string theory is ever to be directly tested, it is probably
necessary to extract some general, qualitative prediction.  One
such prediction might be that there should be low energy
supersymmetry, broken in some particular way.  Another might be
that the string scale is very low, so that there might be many new
states at accessible energies.  In this talk, we have explored
some possibilities, but we do not have firm answers.  Phenomenologically,
a low
string scale is not ruled out, though it may be hard to understand
a scale less than about $6-10 ~{\rm TeV}$. On theoretical grounds,
however, this possibility seems unlikely.  It requires that the
minimum of the potential lie in at a rather implausible extreme
of the
moduli space. It also requires a rather elaborate structure,
and some number of fine tunings.  Still,
absent a real theory, these arguments can at best be described as
informed prejudice.  The challenge for these ideas is to provide
some compelling argument that the scale should, indeed, be at some
particular, low value.  Alternatively, we might be lucky and make
the extraordinary discovery that Newtonian gravity is modified at
short distances, or that phase space is more than four dimensional
at ${\rm TeV}$ energies.

It could be that the usual arguments based on hierarchies for low
energy supersymmetry are incorrect, and that there are good string
ground states in which supersymmetry is badly broken.  After all,
our failure to understand the cosmological constant problem
suggests that our ideas about naturalness and fine tuning are not
entirely correct. So it is  hard, given the present state of our
understanding, to argue persuasively that low energy supersymmetry
is an outcome of string theory.  But we have seen that the
hypothesis of low energy supersymmetry, combined with maximally
enhanced symmetry, makes some definite, qualitative predictions.
We have argued that with these suppositions, supersymmetry must be
broken at very low energy scales (perhaps a few orders of
magnitude above the weak scale) with gauge interactions as the
messengers of the breaking. This suggests, in fact, an approach to
phenomenology which does not require complete control of strong
dynamics.  One might hope to study moduli spaces of $N=1$
theories, and to determine the symmetry structure at their enhanced
symmetry points. Some features, such as spectra and perhaps gauge
couplings and some terms in the superpotential, might be
restricted by symmetries and holomorphy.

It may well be that fundamental theory is entering an era where
hypotheses will be tested principally by their self consistency,
and by considering various gadanken experiments.  But it would be
disappointing if we did not have some picture of how string theory
made contact with nature, and if this picture did not make some
predictions.  It is quite possible that none of the proposals for
string dynamics described here are correct.  The cosmological
constant and the question of the smallness of the gauge couplings
are serious challenges to the maximal symmetry hypothesis, in
particular. But hopefully there is some approach which allows a
qualitative - and perhaps somewhat quantitative - picture.

%\end{document}

\noindent
{\bf Acknowledgements:}

\noindent
I would like to thank T. Banks, A. Nelson, Y. Nir
and Y. Shadmi for enjoyable collaborations and many
discussions of the issues reviewed here.
I would also like to thank S. Dimopoulos and
N. Arkani-Hamid for many patient explanations
of the ideas surrounding large
dimensions.  This work supported in part by the U.S.
Department of Energy.

%%%%%%%%%%%%%%%%%%%%%%%%%%%%%%
%  Bibliography
%%%%%%%%%%%%%%%%%%%%%%%%%%%%%%

\end{document}